\documentclass[aps,superscriptaddress,twocolumn,floats,showpacs,prl,amsmath,amssymb,floatfix,nofootinbib,balancelastpage]{revtex4}

\input epsf
\usepackage{psfig}

\newcommand{\ApJL}{Astrophys. J. Lett.}

\newcommand{\bn}{{\hat{\bf n}}}

\begin{document}

\title{Problems with Pencils: Lensing Covariance of Supernova Distance Measurements}
\author{Asantha Cooray}
\affiliation{Center for Cosmology, Department of Physics and Astronomy, 
  University of California, Irvine, CA 92697}
\author{Dragan Huterer} 
\affiliation{Kavli Institute for Cosmological Physics and Department 
of Astronomy and Astrophysics, University of Chicago, Chicago, IL 60637}
\author{Daniel E. Holz}
\affiliation{Kavli Institute for Cosmological Physics and Department 
of Astronomy and Astrophysics, University of Chicago, Chicago, IL 60637}
\affiliation{Theoretical Division, Los Alamos National Laboratory, Los Alamos, NM 87545}

\begin{abstract}
While luminosity distances from Type Ia supernovae (SNe) provide a powerful
probe of cosmological parameters, the accuracy with which these distances can
be measured is limited by cosmic magnification due to gravitational lensing by
the intervening large-scale structure.  Spatial clustering of foreground mass
fluctuations leads to correlated errors in distance estimates from SNe.  By
including the full covariance matrix of supernova distance measurements, we
show that a future survey covering more than a few square degrees on the sky,
and assuming a total of $\sim$2000 SNe, will be largely unaffected by
covariance noise. ``Pencil beam'' surveys with small fields of view, however,
will be prone to the lensing covariance, leading to potentially significant
degradations in cosmological parameter estimates.  For a survey with 30 arcmin
mean separation between SNe, lensing covariance leads to a $\sim$45\% increase
in the expected errors in dark energy parameters compared to fully neglecting
lensing, and a $\sim$20\% increase compared to including just the lensing
variance. Given that the lensing covariance is cosmology dependent and cannot
be mapped out sufficiently accurately with direct weak lensing observations,
surveys with small mean SN separation must incorporate the effects of lensing
covariance, including its dependence on the cosmological parameters.
\end{abstract}
\pacs{PACS number(s): 98.80.-k, 98.62.Py, 98.80.Es, 97.60.Bw }
\maketitle

{\it Introduction.}---Type Ia supernovae (SNe) have proven to be powerful
probes of the expansion history of the universe~\cite{Rieetal04}, contributing
to the discovery that this expansion is accelerating. A mysterious dark energy
component that comprises $\sim$70\% of the energy density of the universe is
presumed to be responsible for this acceleration.  While the presence
of dark energy is by now well established, its properties and provenance remain
a complete mystery. As the precise nature of the dark energy has profound
implications for both cosmology and particle physics, the elucidation of its
properties is one of the foremost observational and theoretical challenges. It
is hoped that more accurate cosmological measurements will constrain parameters
describing dark energy, and eventually shed light on the underlying physical
mechanism \cite{Huterer}.  Several ongoing programs, including the Supernova Legacy
Survey\footnote{http://www.cfht.hawaii.edu/SNLS/}, Carnegie Supernova
Project\footnote{http://csp1.lco.cl/$\sim$cspuser1/CSP.html},
Essence\footnote{http://www.ctio.noao.edu/$\sim$wsne/}, Sloan Supernova Survey, and
Supernova Factory\footnote{http://snfactory.lbl.gov}, are underway to
observe large samples of low, intermediate, and high-redshift SNe and thereby
obtain $\sim 10\%$ constraints on the equation of state parameter of dark
energy.  Future attempts to measure crucial properties of the dark energy, such
as its time evolution, include a dedicated space-based instrument as part of
the NASA/DOE Joint Dark Energy Mission (JDEM).

\begin{figure*}[t]
\centerline{\psfig{file=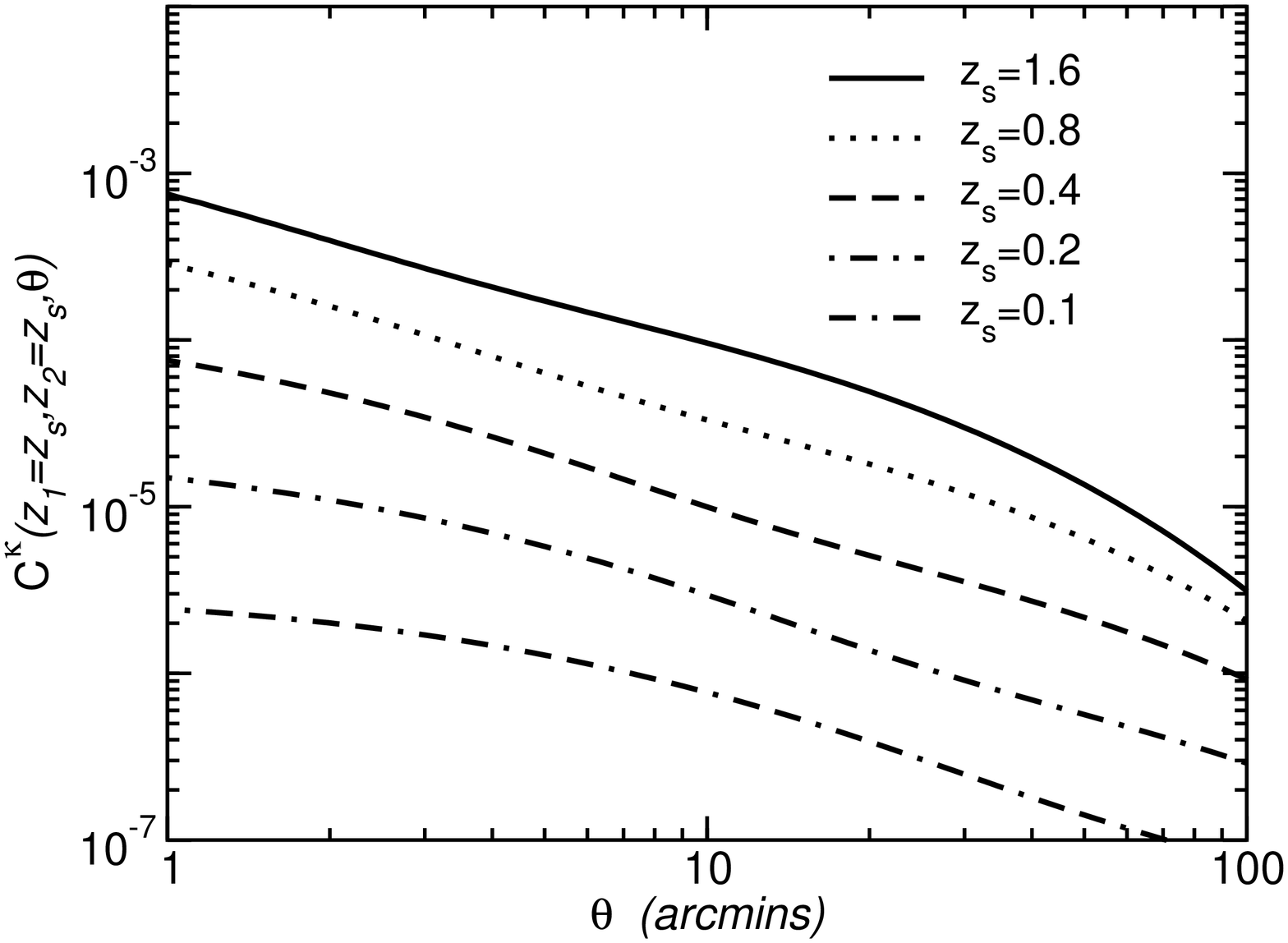,width=4.1in,angle=-90}\hspace{-1cm}
\psfig{file=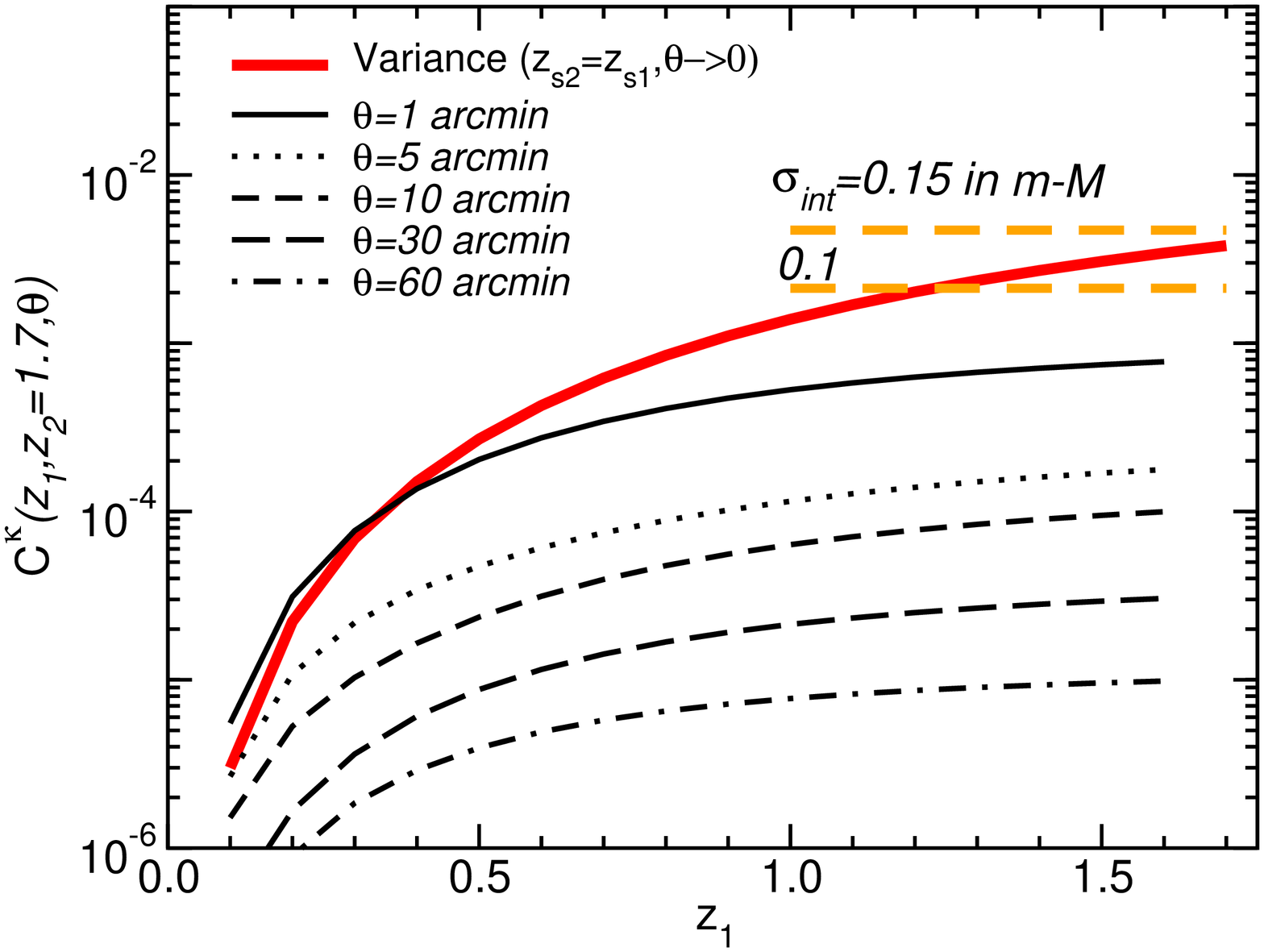,width=4.1in,angle=-90}}\vspace{-0.5cm}
\caption{Covariance of weak lensing convergence, $C^{\kappa}(z_1,z_2,\theta)$,
 as a function of two source redshifts, $z_1$ and $z_2$, and their projected
 angular separation $\theta$. {\it Left panel:} covariance as a function of
 $\theta$ for several values of $z_s\equiv z_1=z_2$.  {\it Right panel:}
 covariance as a function of source redshift with the other source fixed at
 $z=1.7$ and for several illustrative values of $\theta$. For comparison we
 also show the lensing variance as a function of redshift. The two horizontal
 lines represent an intrinsic SN measurement error of $0.10$ and $0.15\,$mag
 (or 0.046 and 0.069 in $\delta d_L/\bar{d}_L$) respectively. Note that lensing
 variance becomes comparable to intrinsic dispersion at $z\gtrsim1.2$ for
 $\sigma_{\rm int}=0.1\,$mag, and $z\gtrsim1.7$ for $\sigma_{\rm
 int}=0.15\,$mag. The lensing variance at low redshift may become smaller than
 covariance of closely separated SNe when one SN is at high redshift.  Of course,
 the correlation coefficient is always less than unity, but can be more than 0.5 if
 the SNe are separated by ten arcminutes or less.}
\label{covar}
\end{figure*}

It is well known that gravitational lensing provides a limit to the accuracy
with which the true luminosity distance can be determined for an individual SN
\cite{Frieman}. The effect comes from the slight modification of the observed
SN flux due to lensing by the intervening large-scale structure. In fact, the
total error budget for SNe at redshifts higher than $z\sim 1$ will have
statistical errors due to lensing comparable to the intrinsic luminosity
distance dispersion \cite{HolzLinder}. These lensing effects may have already
been detected in the current supernova sample \cite{Wang}, although the
evidence is still inconclusive \cite{Menard_Dalal}.  Assuming that lensing
contributes to the variance of the observed SN luminosity distribution (i.e.\
affects each SN observation individually) and using the expected distribution
function for the cosmic magnification \cite{Wamb}, it has been suggested that
the intrinsic power of SNe Ia observation can be restored in the presence of
lensing provided the SN sample is increased by a factor of 2--3
\cite{HolzLinder}.

In addition to the increased variance of SN distance measurements due to
lensing, spatial fluctuations in the foreground mass structures will lead to
correlation of distance estimates of SNe.  Even SNe that are widely separated
in the radial direction will be lensed by common (sufficiently large-scale)
modes of the foreground mass distribution.

In principle, one can use
fluctuations of the mean intrinsic luminosity to measure magnification
statistics \cite{Cooray:05}.  While such measurements are useful in the context
of weak lensing studies, lensing correlations provide a significant challenge
for precision measurement of dark energy properties.  The additional covariance
due to lensing can lead to significant degradation of cosmological parameter
estimates for future small-field SN searches.  It is to be emphasized that our
results apply to any standard-candle approach (e.g., gravitational-wave
standard sirens \cite{HolzHughes}).

\begin{figure*}[t]
\centerline{\psfig{file=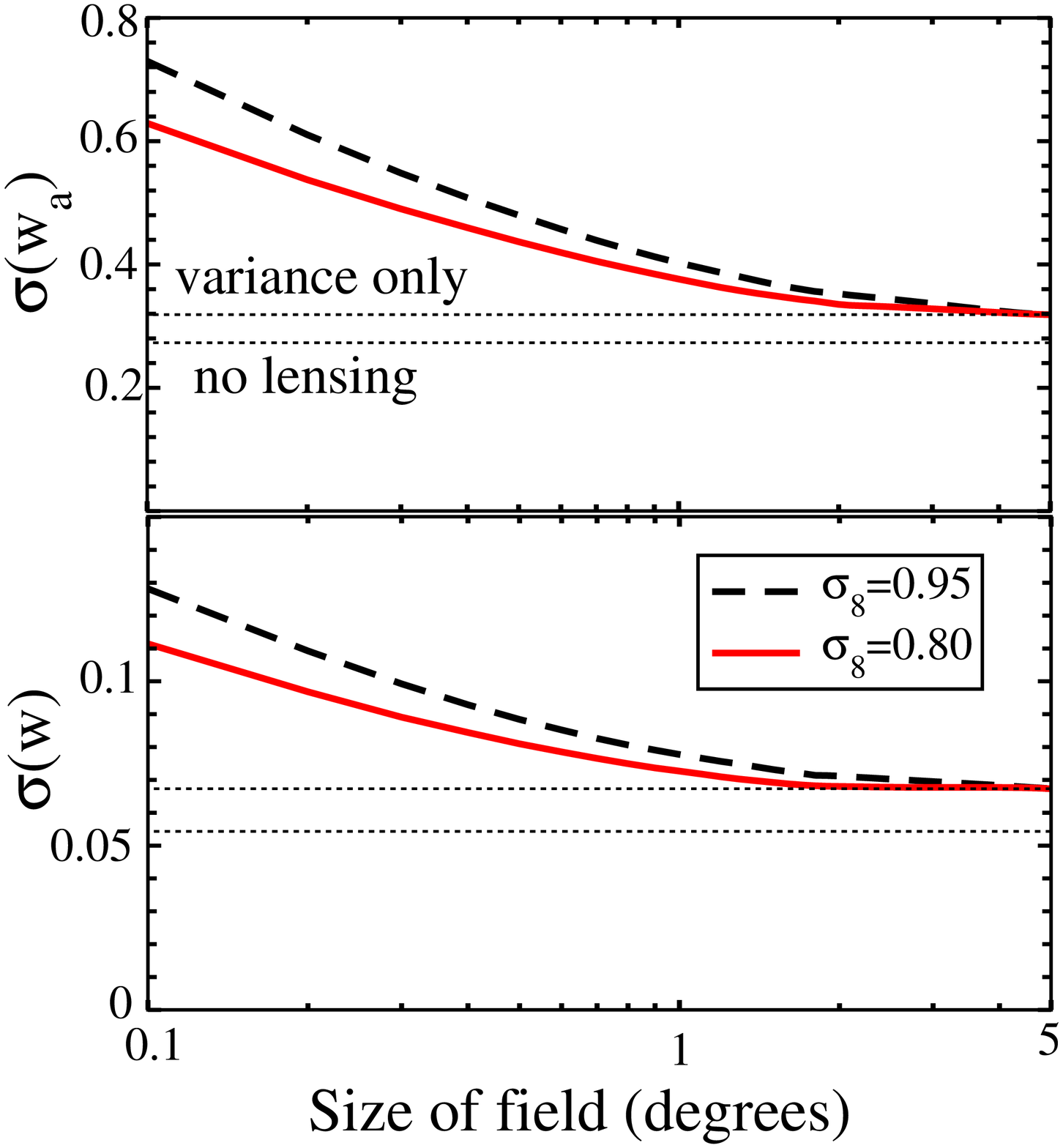,width=3.5in,height=3.8in}
\hspace{0.4cm}
\psfig{file=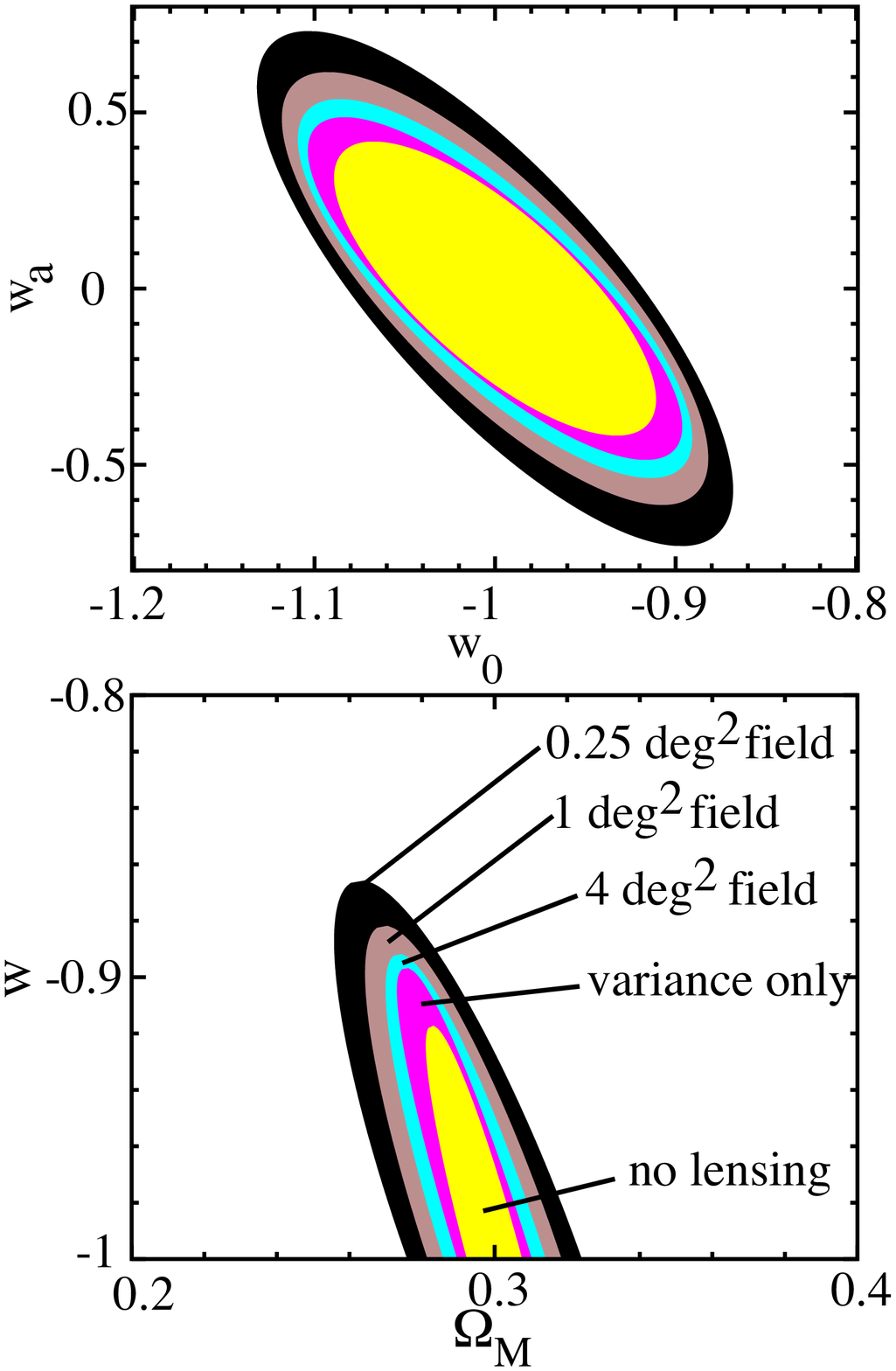,width=3.0in,height=3.8in}}
\caption{{\it Left panel}: Expected errors on $w={\rm const}$ (bottom plot) or
$w_a$ (with a prior on $\Omega_M$ of 0.01; top plot) as a function of the side
length of the observed field. The two dashed curves show errors in
corresponding parameters when lensing is completely ignored, and when only the
lensing variance is considered.  It is apparent that the lensing covariance
contributes to the error budget appreciably when the size of the field is
$\lesssim 1$deg. We show results for two values of $\sigma_8$ that roughly span
the currently favored values of the amplitude of mass fluctuations and hence the SN
lensing covariance.  {\it Right panel:} The full expected constraints projected
into the $\Omega_M-w$ plane (bottom plot; assuming $w={\rm const}$) and
$w_0-w_a$ plane (top plot; with a prior on $\Omega_M$ of 0.01) when
$\sigma_8=0.95$ and for the cases of no lensing, lensing variance only, and a
few selected survey sizes. We have assumed a fixed total number of SNe ($N=1700$)
throughout, regardless of the parameter set and survey sky coverage.}
\label{errors}
\end{figure*}

{\it Calculational Method.}---In order to quantify these statements, we
first summarize the lensing magnification of background SNe due to the foreground
mass distribution and estimate the full covariance matrix associated with
lensing. Lensing modifies the true SN flux by a magnification factor $\mu$,
so that the observed flux is given by $f^{\rm obs}(\bn,z) = \mu(\bn,z) f^{\rm
true}(z)$, where $\bn$ represents the direction of the SN on the sky.
In the weak lensing limit, this magnification can be related to other well-known
quantities through \cite{Bartelmann:1999yn}
\begin{equation}
\mu = [(1-\kappa)^2-|\gamma|^2]^{-1} \approx 1 + 2 \kappa + 3\kappa^2+|\gamma|^2+...\, ,
\label{eq:higher_order}
\end{equation}
where $\kappa (\ll 1)$ is the lensing convergence and
$|\gamma|=\sqrt{\gamma_1^2+\gamma_2^2}$ is the total lensing shear.  Since
$f^{\rm obs} \propto \hat{d}_L^{-2}(z)$, where $d_L(z)$ luminosity distance to
a source at a redshift of $z$, fluctuations in $\mu$ lead to fluctuations in
inferred distance so that $\delta d_L/\bar{d}_L = -\delta \mu$/2. Ignoring
higher order terms (which are suppressed by an order of magnitude in the
lensing variance \cite{Cooray:05,Menard}), one can take $\mu \approx 1 +
2\kappa$ and relate SN distance fluctuations due to lensing to the convergence
along the line-of-sight. Thus, the full covariance matrix of fractional
distance estimates for a sample of supernovae is
\begin{equation}
{\rm Cov}_{ij} \approx  \sigma_{\rm int}^2 \delta_{ij} +
C^{\kappa}(z_i,z_j,\theta_{ij}) \, ,
\label{eq:covar}
\end{equation}
where 
$\sigma_{\rm int}$ is the intrinsic error that affects each SN distance 
measurement.

Using the angular cross power spectrum of convergence between two different
redshifts, computed under the Limber approximation \cite{Kaiser:1996tp}
\begin{eqnarray}
&&C_\ell^{\kappa\kappa}(z_i,z_j) = \nonumber \\
&&\int_0^{{\rm min}(r_i,r_j)} 
dr \frac{W(r,r_i)W(r,r_j)}{d_A^2} P_{\rm dm}\left(k=\frac{l}{d_A(r)},r\right)
\nonumber \\[0.1cm]
&&W(r,r_s) = {3 \over 2} \Omega_m \frac{H_0^2}{c^2 a(r)} 
\frac{d_A(r) d_A(r_s-r)}{d_A(r_s)} \, ,
\label{eq:cl}
\end{eqnarray}
the lensing contribution to the covariance is
\begin{equation}
C^{\kappa}(z_i,z_j,\theta_{ij}) = \int \frac{d^2 {\bf l}}{(2\pi)^2}
C_\ell^{\kappa \kappa}(z_i,z_j) J_0(l\theta_{ij}) \, .
\label{eq:ctheta}
\end{equation}
Here $J_0$ is the 0th order Bessel function of the first kind.  In
Eq.~(\ref{eq:cl}), $r_i$ and $r_j$ are comoving distances corresponding to SNe
at redshifts $z_i$ and $z_j$ respectively, $d_A$ is the angular diameter
distance, and $P_{\rm dm}(k, r)$ is the three-dimensional power spectrum of
dark matter evaluated at the distance $r$; we calculate it using the halo model
of the large-scale structure mass distribution \citep{Cooray:02}.

Eq.~(\ref{eq:covar}) defines the full covariance matrix due to lensing for
supernovae at redshifts $z_i$ and $z_j$ with projected angular separation of
$\theta_{ij}$ on the sky. For reference, the previously considered excess
variance due to lensing corresponds to diagonal elements of ${\rm
Cov}_{ij}$ with $z_i=z_j$ and $\theta_{ij}=0$. In this limit, 
 $J_0(l\theta_{ij}) \rightarrow 1$ in Eq.~(\ref{eq:ctheta}) and one recovers the
variance, $\sigma^2(z) =\int l dl C_\ell^{\kappa \kappa}(z)/2\pi$.

In the left panel of Figure~1 we show the covariance $C^{\kappa}(z_i,z_j,\theta)$ 
as a function of $\theta\equiv \theta_{ij}$ (which is assumed fixed for the moment)
for several values of 
$z_i=z_j$, while in the right panel we show the covariance as a function of $z_1$ with 
the other redshift fixed at $z_2=1.7$. For reference we also plot the variance as a
function of redshift $z$ and compare it to the intrinsic SN magnitude errors
of 0.10 and 0.15 mag, roughly spanning the error expected in upcoming surveys.
To estimate the resulting effect on cosmological parameter estimates, we
compute the Fisher information matrix
\begin{equation}
{\bf F}_{\alpha\beta} = \sum_{ij}
        {\partial d_L(z_i) \over \partial p_\alpha} \left({{\rm Cov}^{-1}}\right)_{ij}
{\partial d_L(z_j) \over \partial p_\beta} \, .
\label{eqn:fisher}
\end{equation}
If the variance of SN distance measurements alone is considered, the Fisher
matrix reduces to the familiar form, with the factors $N(z_i)/(\sigma^2_{\rm
int}+\sigma^2_{\rm lens})$ representing the inverse covariance terms; here
$N(z_i)$ is the number of SNe in the redshift bin centered at $z_i$ and
$\sigma^2_{\rm lens}$ is the variance due to lensing.  With the full covariance
matrix considered, this simple form no longer holds.  Moreover, a full $N_{\rm
tot}\times N_{\rm tot}$ Fisher matrix (and not the redshift-binned smaller
version) is now required in order to obtain the cosmological parameter accuracy
estimates; however, this is not a novel problem since a correct treatment of SN
calibration uncertainties similarly requires the full $N_{\rm tot}\times N_{\rm
tot}$ (or even larger) covariance matrix \cite{Kim_Miquel}.  Here we implicitly
neglect information from the cosmological parameter dependence of the
covariance matrix; 
there would only be significant information in the covariance 
if the off-diagonal terms were comparable to the diagonal ones.

{\it Discussion.}---To estimate cosmological parameter measurement errors, we
assume a survey with 1700 SNe distributed uniformly in redshift out to $z=1.7$
(roughly following Ref.~\cite{SNAP}).  To speed up the calculation of the
$1700\times 1700$ covariance matrix, we compute it in discrete redshift bins,
stepping by 0.1 in both $z_i$ and $z_j$. The covariance also depends on the
angular separation of SNe, and we distribute the SNe randomly in a square field
whose side (or total area) we are free to change. The histogram of the angular
separations is a smooth bell curve that peaks at roughly half the field size.

Figure~2 summarizes the effect of lensing covariance on dark energy
measurements from the assumed future SN survey. We model the evolution of the
dark energy equation of state with redshift as $w(a)=w_0+(1-a)w_a$ where $a$ is
the scale factor, and consider measurements of four parameters: the matter
energy density relative to critical, $\Omega_M$, $w_0$, $w_a$, and the nuisance
parameter $\mathcal{M}$ that combines the Hubble constant and absolute SN
magnitude information. Our fiducial model is standard $\Lambda$CDM with
$\Omega_M=0.3$, $w_0=-1$, and $w_a=0$.  Figure~2 (left panel) shows the
expected errors on $w={\rm const}$ (bottom plot) or $w_a$ (with a prior on
$\Omega_M$ of 0.01; top plot) as a function of the size of the observed
field. The two dotted curves show errors in corresponding parameters when
lensing is completely ignored, and when solely the lensing variance is
considered.  It is apparent that the lensing covariance contributes to the
error budget appreciably when the size of the field is $\lesssim 1$
deg. Furthermore, the effects of lensing covariance depend on the fiducial
convergence power $C^{\kappa}(z_i,z_j,\theta_{ij})$, which in turn is sensitive
to the amplitude of mass fluctuations $\sigma_8$ (and, to a lesser extent,
other cosmological parameters). Since $\sigma_8$ is somewhat uncertain at
present, we show results for two values, $\sigma_8=0.8$ and $\sigma_8=0.95$,
that roughly span the currently favored values of the amplitude of mass
fluctuations in the universe.

Figure~2 (right panel) shows the full expected constraints projected into the
$\Omega_M-w$ plane (bottom plot; assuming $w={\rm const}$) and $w_0-w_a$ plane
(top plot; with a prior on $\Omega_M$ of 0.01) with $\sigma_8=0.95$ and for the
cases of no lensing, lensing variance only, and a few selected survey
sizes. Again we see that surveys of less than about one square degree will
suffer from considerable error due to lensing covariance.  As the mean
separation between supernovae is increased, off-diagonal terms in the
covariance matrix decrease, and the resulting effect on cosmological parameters
is reduced.

Our results can be understood simply in the limit of equal off-diagonal
covariance terms. In this case, the Fisher matrix estimate of error in
parameter $p_\alpha$ is increased by a factor $\sqrt{1+(N-1)r^2}$ relative to
the case with no off-diagonal terms, where
$r=C^\kappa(z_i,z_j,\theta)/\sqrt{C^\kappa(z_i,z_i,0)C^\kappa(z_j,z_j,0)}$ and
$N$ is the total number of SNe in the sample. With $N \sim$ 2000 or more in
upcoming searches, parameter errors will increase by a factor of $\sqrt{2}$
when $r \sim 1/\sqrt{N} \sim 0.02$. Furthermore, Figure~1 reveals that when
$z_1\gtrsim 1$ correlations are at the percent level when SNe are separated by
$\theta \sim 10$ arcmin.  Note that in order to accurately estimate the errors
on dark energy parameters one will need to allow for the dependence of the
covariance matrix on imprecisely known cosmological parameters that determine
the weak lensing convergence power spectrum. Accounting for the uncertainty in
the covariance is even more important since galaxy shear maps are not useful to
correct for lensing-modified SN fluxes \cite{Dalal}.

{\it Conclusions.}---We have discussed gravitational lensing covariance as an
additional source of error for cosmological surveys utilizing standard candles.
Future supernova surveys that plan $\sim 10$--$20$ deg$^2$ coverage with $\sim
\mbox{2000}$ SNe will be largely unaffected by lensing covariance. Lensing
variance remains an issue, but is reduced through increased numbers of SNe
(about 50 SNe per redshift bin of width 0.1 are necessary to reduce the lensing
variance so that it is negligible compared to the systematic floor
\cite{HolzLinder}). The cosmological parameter accuracies for a survey with a
rectangular field that is wide in one direction and narrow in another may be
compromised, since the histogram of the angular distribution of SNe now has a
peak at an angle of order the narrow side of the survey (albeit with a very
pronounced tail).  We find that true ``pencil beam'' surveys with a sky
coverage of $\lesssim 1$ deg$^2$ in a single field are subject to significant
degradation in cosmological parameter accuracies due to lensing covariance.
The consideration of lensing covariance thus argues against pencils, and in
favor of wider-field surveys.

\begin{acknowledgments}
We thank Alex Kim for useful comments on the manuscript.  D.H.\ is supported by
an NSF Astronomy and Astrophysics Postdoctoral Fellowship under Grant No.\
0401066. D.E.H. acknowledges a Richard P. Feynman Fellowship from Los Alamos
National Laboratory.
\end{acknowledgments}

\end{document}